\documentclass[epj]{svjour}
\usepackage{graphics}
\usepackage{graphicx}
\usepackage[dvipsnames,rgb]{xcolor}
\usepackage{dcolumn}
\usepackage{psfrag}
\usepackage{float,color}
\usepackage{textcomp}
\usepackage{units}
\usepackage{hyperref}
\usepackage{amsmath}
\usepackage{amssymb}
\usepackage{amsopn}

\newcommand{\marker}[1]{\protect\includegraphics[width=2mm,clip]{mark#1.pdf}}

\newcommand{\ve}[1]{\ensuremath{\mbox{\boldmath$#1$}}}
\newcommand{\sve}[1]{\ensuremath{\mbox{\footnotesize\boldmath$#1$}}}
\newcommand{\ma}[1]{\ensuremath{\mathbb{#1}}}

\DeclareMathOperator{\ku}{Ku}
\DeclareMathOperator{\st}{St}
\DeclareMathOperator{\J}{{\cal J}}
\DeclareMathOperator{\R}{{\cal R}{}}
\DeclareMathOperator{\Rt}{\widetilde{\cal R}{}}

\newcommand{\eqnlab}[1]{\label{eq:#1}}
\newcommand{\figlab}[1]{\label{fig:#1}}

\newcommand{\eqnref}[1]{(\ref{eq:#1})}
\newcommand{\figref}[1]{\ref{fig:#1}}

\newcommand{\Eqnref}[1]{Eq.~(\ref{eq:#1})}
\newcommand{\Figref}[1]{Fig.~\ref{fig:#1}}

\newcommand{\seclab}[1]{\label{sec:#1}}

\newcommand{\Secref}[1]{Section~\ref{sec:#1}}

\begin{document}
\title{Statistical model for collisions and recollisions of inertial particles in mixing flows$^\star$}
\author{K. Gustavsson\inst{1}\inst{2} \and B. Mehlig\inst{1}
}                     
%
%
\institute{Department of Physics, Gothenburg University, 41296 Gothenburg, Sweden \and Department of Physics and INFN, University of Rome 'Tor Vergata', Via della Ricerca Scientifica 1, 00133 Rome, Italy}
\date{{$^\star$} Version accepted for publication (postprint) in The European Physical Journal E {\bf 39}, 55 (2016)}
%
\abstract{
Finding a quantitative description of the rate of collisions between small particles suspended in mixing flows is a long-standing problem.  Here we investigate the validity of a parameterisation of the collision rate for identical particles subject to Stokes force, based on results for relative velocities of heavy particles that were recently obtained within a statistical model for the dynamics of turbulent aerosols. This model represents the turbulent velocity fluctuations by Gaussian random functions. We find that the parameterisation gives quantitatively good results in the limit where the \lq ghost-particle approximation' applies.  The collision rate is a sum of two contributions due to \lq caustics' and to \lq clustering'. Within the statistical model we compare the relative importance of these two collision mechanisms.  The caustic formation rate is high when the particle inertia becomes large, and we find that caustics dominate the collision rate as soon as they form frequently.  We compare the magnitude of the caustic contribution to the collision rate to the formation rate of caustics.
} 
\maketitle
%
%
%
%

\section{Introduction}
\seclab{introduction}
To understand how particles in turbulent flows aggregate requires a quantitative model for
their collision rate.
This is important for a wide range of problems. The growth of rain droplets settling
in turbulent cumulus clouds \cite{Dev12,Bod10}, and planet formation in circumstellar
accretion discs~\cite{Arm07,Joh14} are two examples where turbulence may affect aggregation.
But to derive a quantitative parameterisation of the collision rate of particles in turbulent aerosols from first principles
is a very complicated problem.  It is often simplified by assuming that the particles are
spherical and very small, and that they do not directly
interact with each other.
Commonly, mono-disperse suspensions of particles are considered (this is the simplest case), and the so-called
\lq ghost-particle approximation\rq{} is employed where the particles are allowed
to move through each other upon collision \cite{Zho98,Vos13}.
\lq Collisions\rq{} are counted in this picture by recording how frequently particles approach
closer than their collision radius (twice the particle radius).
A highly idealised model, frequently adopted in the literature retains only Stokes force.
This yields the equation of motion
\begin{equation}
\dot{\ve r}=\ve v\,,\hspace{0.5cm}
\dot{\ve v}=\gamma(\ve u(\ve r,t)-\ve v)\,.
\eqnlab{eqm_Stokes}
\end{equation}
Here dots denote time derivatives, $\ve r$ is the position of a particle, $\ve v$ is its velocity and $\ve u$ is the fluid velocity evaluated at the particle position.
Further $\gamma$ is the Stokes damping rate of the particle. The \lq Stokes number' $\st=1/(\gamma\tau)$ is a dimensionless measure
of the particle inertia, $\tau$ is the smallest time scale of the flow.

A number of different parameterisations for the collision rate have been
suggested in the literature, attempting to describe the effect of particle inertia upon the collision rate,
see e.g. Refs.~\cite{Sun97,Fal02,Wil06}.
A recent theory for the relative velocities of particles in turbulence \cite{Gus11b,Gus12,Gus13c} is based on a statistical model for the
turbulent fluctuations in the dissipative range, it gives a quantitative description of DNS results for the moments of relative velocities at
small separations \cite{Bec11} and for the corresponding distribution provided that the Stokes number is not so large that inertial-range fluctuations
contribute substantially \cite{Per15}.  This theory yields an expression for the collision rate (in the ghost-particle approximation), and
allows to quantify the relative importance of two important mechanisms for collisions: \lq clustering' and \lq caustics'.
Spatial clustering of particles on a fractal attractor increases the probability to find close-by particles and this may enhance the rate of collisions~\cite{Sun97}.
The mechanisms for fractal small-scale clustering of heavy particles in turbulence are reviewed in \cite{Gus15a}.
Caustics occur as the phase-space manifold folds over, allowing particles coming from far apart to collide at small separations with large relative velocities~\cite{Fal02,Wil06}.
The results in Refs.~\cite{Gus11b,Gus12,Gus13c} provide a simple parameterisation of the two contributions to the collision rate, allowing
us to quantify the relative importance of the contributions due to caustics and clustering.

In this paper we present a summary of the derivation of the parameterisation in Refs.~\cite{Gus11b,Gus12,Gus13c}.
We explicitly compare the two contributions to simulations of the collision rate obtained for a statistical model that represents the turbulent velocity field by a Gaussian random function.

We outline the derivation of an estimate, first derived in Ref.~\cite{Gus13c}, for the contribution to the collision rate due to clustering in the statistical model.
Subtraction of this contribution from the full collision rate allows us to isolate the contribution due to caustics.

We find that the contribution to the collision rate due to caustics is
consistent with results of direct numerical simulations~\cite{Fal07c,Duc09}, where it was found that it can be responsible for up to 50\% of the total collision rate for $\st$ of order unity.  It is also consistent with the results in Ref.~\cite{Vos13}, where it was found that caustics dominate the collision rate for $\st>0.3$.

As shown in Refs.~\cite{Bru98,And07,Gus08,Vos11,Pec12}, the ghost-particle approximation fails in the advective limit due to recollisions between close-by ghost particles.
We investigate how large this effect is for finite Stokes numbers and for different particle sizes.
Our results are consistent with the numerical results obtained in Ref.~\cite{Vos11}.

\section{Statistical model}
\seclab{model}
We take $\ve u$ in \Eqnref{eqm_Stokes} to be an isotropic, homogeneous, incompressible, and Gaussian random velocity field~\cite{Wil06,Gus13c,Gus15a}.
We use a single-scale flow: $\ve u$ has a single length scale $\eta$, time scale $\tau$, and speed scale $u_0$.
Our numerical simulations are carried out in $d=2$ spatial dimensions where we can write $\ve u=u_0\ve\nabla\phi\wedge\hat{\ve e}_3/\sqrt{2}$, where $\hat{\ve e}_3$ is a unit vector in the direction orthogonal to the simulation plane.
The stream function $\phi$ is a Gaussian random function with zero mean and correlation function
\begin{equation}
\langle\phi(\ve x,t)\phi(\ve 0,0)\rangle={\rm e}^{-|\sve x|^2/(2\eta^2)-|t|/\tau}\,.
\eqnlab{phiCorr}
\end{equation}
Here $\langle X\rangle$ denotes an ensemble average of $X$.
In addition to St, a second dimensionless parameter is the \lq Kubo number', $\ku=u_0\tau/\eta$, which is of order unity or larger in turbulent flows~\cite{Gus15a}.
In the following we assume that the particle radius $a$ is small, $a/\eta \ll 1 $, and we vary the dimensionless
parameters $a/\eta$ and $\st$ independently.
We also assume that the fraction of total droplet volume compared to the system volume is small to ensure that backreaction on the fluid is negligible.

\section{Parameterisations of the collision rate}
\seclab{parameterisation}
When $\st=0$ particles are brought into contact by smooth shearing of fluid elements.
The collision rate of particles in a constant simple shear was found analytically by Smoluchowski~\cite{Smo17} to scale as $\sim s n_0a^d$ per particle, where $s$ is the shear rate, $n_0$ the number density of particles, and $d$ the spatial dimension.
In general a flow with a constant strain matrix, $\ma S\equiv(\ve\nabla\ve u^{\rm T}+(\ve\nabla\ve u^{\rm T})^{\rm T})/2$, gives a collision rate that scales as $\sim n_0a^d$. The prefactor depends on the eigenvalues of the strain matrix~\cite{Gus08}.
Saffman and Turner~\cite{Saf56} argued that since advected particles are uniformly distributed in an incompressible flow, their collision rate is given by an ensemble average over strain matrices $\ma S$ using the distribution of $\ma S$ in the flow.
The corresponding collision rate is $\R_{\rm ST}\sim n_0a^d$ with a flow-dependent prefactor.
For the statistical model in \Secref{model} this rate was evaluated analytically in Ref.~\cite{Gus08}
\begin{align}
\R_{\rm ST}\tau=\sqrt{\frac{d}{2\pi}}N\ku\left(\frac{2a}{L}\right)^d\,,
\eqnlab{R_ST}
\end{align}
where $N$ is the total number of suspended particles in a large sphere of radius $L\gg\eta$.

It was observed by Brunk, Koch and Lion \cite{Bru98} that the Saffman-Turner expression $\R_{\rm ST}$ overestimates the collision rate in flows with a rotational component because each recollision between two close-by particles gives a contribution to $\R_{\rm ST}$.
In Refs.~\cite{And07,Gus08} this effect was evaluated analytically for particles advected in rapidly fluctuating velocity fields, in the limit of small Kubo numbers.
When $\ku$ is small the separation between two particles undergoes a diffusion process.
The probability distribution for particle separations is found by solving a Fokker-Planck equation with absorbing boundary conditions at the collision distance $2a$.
From this distribution the ingoing probability flux evaluated at the collision distance yields the collision rate~\cite{Gus08}
\begin{align}
\R\tau=\frac{d}{d-1}N\ku^2\left(\frac{2a}{L}\right)^{d}\,.
\eqnlab{R_advcol}
\end{align}
This expression is valid for small values of $\ku$ and $\st$ for the statistical model described in \Secref{model}.

A second limit in which the collision rate has been estimated is that of large values of $\st$. In this limit Abrahamson estimated the collision rate using kinetic-gas theory~\cite{Abr75}.
In the limit of large $\st$ the velocity field $\ve u$ fluctuates rapidly compared to the particle response time. Consequently the particle velocity undergoes an Ornstein-Uhlenbeck process.
It follows that particle velocities are Gaussian and the collision rate was estimated as $\R_g\tau\sim \sqrt{\langle V_R^2\rangle}a^{d-1}$~\cite{Abr75}, where $V_R$ is the radial velocity between two particles.
For the statistical model the gas-kinetic collision rate evaluates to
\begin{align}
\R_g\tau=\sqrt{\frac{d}{\pi\st}}N\ku\frac{\eta}{L}\left(\frac{2a}{L}\right)^{d-1}\,.
\eqnlab{R_KG}
\end{align}
The size-dependence in $\R_g\sim a^{d-1}$ differs from that found at small values of $\st$, $\R\sim a^d$.
The different scalings are explained by the fact that when $\st=0$ the smooth dynamics between two close-by particles leads to typical collision velocities of the order of $~u_0a/\eta$ upon contact, while in the kinetic-gas limit collision events are decoupled from the instantaneous local fluid velocity (due to the formation of caustics), leading to collision velocities $\sim u_0$.
The additional factor $a^{d-1}$ in the expressions for the collision rates is simply a geometric contribution from the spatial volume element at the collision distance $2a$.

A number of parameterisations for the collision rate at intermediate values of $\st$ have been suggested.
In Refs.~\cite{Fal02,Wil06,Duc09,Vos13} it is argued that the collision rate of inertial particles is the sum of two contributions, due to clustering and due to caustics.
In Refs.~\cite{Sun97,Bec05,Chu05}, by contrast, the collision rate is parameterised as a product of two factors.
The relation between these different parameterisations is discussed in \Secref{collision_smooth}, see also Ref.~\cite{Gus13c}.
Different parameterisations emphasise different aspects of the problem (clustering, caustics).

As an example, in Ref.~\cite{Wil06} the collision rate of inertial particles was parameterised as
\begin{equation}
\R\sim \R_{\rm ST}+\exp(-{S}/\st)\R_{\rm g}\,.
\eqnlab{MWB_ansatz}
\end{equation}
The $\st$-dependent prefactor of $\R_{\rm g}$ determines the relative weights of the two terms.
In Ref.~\cite{Wil06} it was taken to be proportional to the rate of caustic formation in a one-dimensional white-noise model for inertial particles for small values of $\st$, $\propto \exp(-S/\st)$.
This was motivated by the expectation that the number of collision events due to caustics is closely related to the rate at which caustics form.
This ansatz leads to the prediction that the second term in \Eqnref{MWB_ansatz} dominates at large values of $\st$ for small particles.
But~\eqnref{MWB_ansatz} does not allow to quantify the relative importance of clustering and caustics.
The \lq action' $S$ in \Eqnref{MWB_ansatz} is a system-dependent number of order unity. In Ref.~\cite{Wil06} it was fitted to numerical data, see also Ref.~\cite{Fal07c}.

\section{Collision rate and relative velocities}
\seclab{collrate_relvel}

In the following Sections we use numerical simulations of the statistical model to investigate the validity of the parameterisation for the collision rate given by Eqs.~\eqnref{Rt0} and~\eqnref{collision_relvel_moment1} below.
\Eqnref{collision_relvel_moment1} was first derived in the white-noise limit in Ref.~\cite{Gus11b}, and at finite Kubo numbers in Ref.~\cite{Gus13c}.

The collision rate in the ghost-particle approximation, $\Rt$,
is closely related to the first moment of relative velocities between
the suspended particles evaluated at the collision radius,
\begin{eqnarray}
\label{eq:Rt0}
\Rt &=&  N\int{\rm d}V_R\,|V_R|\,\rho(2a,V_R)\Theta(-V_R)\nonumber\\
&\approx &\frac{N}{2} \int{\rm d}V_R\,|V_R|\,\rho(2a,V_R) \\
&\equiv& \frac{N}{2}  m_1(2a)\,.\nonumber
\end{eqnarray}
As before $N$ is the total number of suspended particles, $2a$ is the distance below which particles collide,
and $\rho(R,V_R)$ is the probability for two particles to be at a distance $R$ with relative radial velocity $V_R=\Delta\ve v\cdot\hat{\ve e}_R$,
where $\hat{\ve e}_R$ is the unit vector along the separation between the two particles and $\Delta\ve v$ is the relative velocity of the two particles.
The probability distribution $\rho(R,V_R)$ is normalised to unity over a large sphere of radius $L\gg\eta$.
The properties of this joint distribution was studied in Refs.~\cite{Gus11b,Gus12,Gus13c}.

In Ref.~\cite{Gus13c} the moment $m_1(R)$ needed to estimate the ghost-particle approximation in \Eqnref{Rt0} was derived as follows.

First, two asymptotic limits of the joint distribution $\rho(\Delta\ve v,\Delta\ve r)$ of relative velocities $\Delta\ve v$ and separations $\Delta\ve r$ for a particle pair are matched.
The first limit assumes that caustics may occur in the system, as is the case for any non-zero value of $\st$ in the dynamics of inertial particles.
Caustics give rise to isotropic uniform motion at small enough distances, ${\gamma}R\ll |\Delta\ve v|\equiv V$, leading to $\rho(\Delta\ve v,\Delta\ve r)$ being a function of $V$ only in this limit.
In the second limit, ${\gamma}R\gg V$, particle separations undergo slow isotropic diffusion due to variations of the small $\Delta\ve v$ and the distribution is approximated by a function of distance $R$ only.

The asymptotic behaviours of the distribution in these two limits are determined by matching their functional forms along a curve $V\propto\gamma R$ and by using the observation that inertial particles show fractal clustering in phase-space.
This matching curve is appropriate when colliding particles originate from smooth dynamics.
For large enough values of $\st$ in turbulent flows this matching curve may need to be modified to take into account collisions between particles originating from the inertial range.

If particles show fractal clustering in phase-space, the distribution $\rho(w)$ of small phase-space separations $w\equiv\sqrt{R^2+(V/\gamma)^2}$ has a power-law dependence, $\rho(w\ll 1)\sim w^{D_2-1}$.
Here $D_2$ is the \lq phase-space correlation dimension' of the fractal attractor.
The power-law dependence of $\rho(w)$ is used as a boundary condition to determine the asymptotic approximate form of $\rho(\Delta\ve v,\Delta\ve r)$.
The resulting asymptotic distribution is built from joined power-laws in $R$ and $V$. The the powers are determined by the spatial dimension $d$ and by $D_2$.

Given the form of the distribution one can determine the moment $m_1(R)$ in \Eqnref{Rt0}.
For small values of $R$ it takes the form~\cite{Gus11b,Gus13c}
\begin{equation}
m_1(R)\sim \underbrace{b_1 (R/\eta)^{D_2}}_{\mbox{smooth}}+\underbrace{c_1 (R/\eta)^{d-1}}_{\mbox{caustic}}\,.
\eqnlab{collision_relvel_moment1}
\end{equation}
Eqs. (\ref{eq:collision_relvel_moment1}) and \eqnref{Rt0} yield an expression for the collision rate that is the sum of two contributions, just as \Eqnref{MWB_ansatz}.
An important difference to \Eqnref{MWB_ansatz} is that the parameterisation~\eqnref{collision_relvel_moment1} takes into account particle clustering. A second difference is that the caustic term is weighted by the coefficient $c_1$ which characterises the contribution of caustics to the moment $m_1$ of relative radial velocities [in contrast to the caustic formation rate in~\eqnref{MWB_ansatz}].
The parameterisation \eqnref{collision_relvel_moment1} allows to quantify the relative importance of clustering and caustics upon the collision rate.
Eq.~(\ref{eq:collision_relvel_moment1}) is closely related to the theory put forward in Ref.~\cite{Vos13}.

Since little was assumed concerning the flow in deriving \Eqnref{collision_relvel_moment1}, we expect it to be valid with great generality.
We remark that it may be necessary to refine the theory if the inertial range becomes important at large Stokes numbers.
It must also be noted that the parameters in \Eqnref{collision_relvel_moment1} are non-universal, they depend upon the flow and on the Stokes number.
In the following we discuss the system-dependent parameters $b_1$ and $c_1$ in \Eqnref{collision_relvel_moment1} for the statistical model described in \Secref{model}.

\begin{figure*}
\includegraphics[width=18cm]{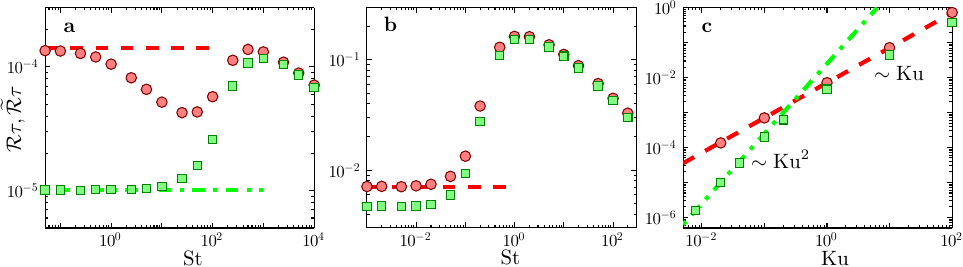}
\caption{\figlab{advective} {\em (Online colour).}
Markers show data from numerical simulations of the collision rate
$\R$ (\marker{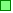}) and its ghost-particle approximation $\Rt$ (\marker{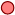}).
Panels {\bf a} and {\bf b} show $\R$ and $\Rt$ as functions of $\st$ for $\ku=0.02$ ({\bf a}) and $\ku=1$ ({\bf b}).
Panel {\bf c} shows $\R$ and $\Rt$ as functions of $\ku$ for $\st=0$.
Red dashed lines show the Saffman-Turner estimate $\R_{\rm ST}$ \eqnref{R_ST} for the ghost-particle approximation, and green dash-dotted lines show the small-$\ku$ theory for the collision rate \eqnref{R_advcol}.
Parameters: $a=0.01\eta$, $N=1000$, $L\approx 5\eta$.
}
\end{figure*}

\section{Recording collisions and recollisions}
We have performed a number of simulations of $N$ particles moving independently according to \eqnref{eqm_Stokes} in a two-dimensional incompressible Gaussian random
velocity field $\ve u$. A collision occurs when the separation between two particles becomes less than the collision radius, $2a$.
To calculate the collision rate $\R$ we remove one of the colliding particles from the system after the collision to avoid recollisions~\cite{And07,Gus08,Vos11}.
We expect this to be a valid approach for dilute systems with high collision- and coalescence efficiency.
In the ghost-particle approximation the particles are allowed to recollide as explained in the introduction.
We denote the corresponding collision rate by $\Rt$.
In this case particles may continue to collide frequently after their first collision.
In numerical simulations we calculate the steady-state collision rate by counting the total number of collisions during a time interval (neglecting initial transients) and divide this number by the length of the time interval.
\Eqnref{Rt0} is the collision rate per particle in the ghost-particle approximation,
the pairwise collision rate per particle pair is given by multiplying \eqnref{Rt0} by $(n_0-1)/2$ where $n_0$  is the number density of the particles.
When a particle is removed upon collision we adjust the collision rate to compensate for the decrease in the number of particles.

Figs.~\figref{advective}{\bf a} and {\bf b} compare the collision rate with its ghost-particle approximation as a function of $\st$ for two different values of $\ku$. The data is taken from Ref.~\cite{Gus11c}. We see that the ghost-particle approximation agrees with the Saffman-Turner estimate [\Eqnref{R_ST}], but deviates from the collision rate $\R$ for small values of $\st$.
This fact ($\R\ne\Rt$ at $\st=0$) is clearly visible in \Figref{advective}{\bf a} and {\bf b}.
For large enough values of $\st$ the two estimates approximately agree.
Similar behaviour is observed in kinematic simulations of particles suspended in turbulent flows~\cite{Vos11}.

The expression for the collision rate for small values of $\ku$ and $\st=0$, \Eqnref{R_advcol}, is shown in \Figref{advective}{\bf a} and {\bf c}.
\Figref{advective}{\bf c} shows that it scales as $\ku^2$ [\Eqnref{R_advcol}], while the Saffman-Turner rate scales as $\ku$ [\Eqnref{R_ST}]. For larger values of $\ku$ the Saffman-Turner estimate is an upper bound on the collision rate.

\section{Smooth contribution}
\seclab{collision_smooth}
The smooth contribution to the collision rate [first term in \eqnref{collision_relvel_moment1}] can be estimated for general values of $\st$ in the ghost-particle approximation by using
earlier results for $m_1$.
These results were derived assuming that relative radial velocities are close to Gaussian and that instantaneous correlations between the particle positions and the structures of the flow, so called \lq preferential sampling'~\cite{Gus15a}, can be neglected.
The non-Gaussian nature of turbulent flows, as well as the possible existence of persistent flow structures that are preferentially sampled may make it necessary to modify this estimate for turbulent flows.
Following the derivation in Ref.~\cite{Gus13c} we obtain an estimate for $m_1(R)$ needed for the collision rate as follows.
First, we write

\begin{equation}
m_1(R)=m_0(R)S_1(R)\,.
\eqnlab{mp_smooth}
\end{equation}
Here $m_0(R)$ is the probability distribution of $R$, and the structure function $S_1(R)$ is the average of $|V_R|$ conditional on $R$~\cite{Gus13c}.
The form \eqnref{mp_smooth} is reminiscent of the parameterisation of the collision rate in terms of a product of two contributions suggested in Refs.~\cite{Sun97,Bec05,Chu05}.
The moment $m_0(R)$ is related to the pair correlation function $g(R)$ as $m_0(R)\sim g(R)R^{d-1}$~\cite{Sun97}.
But since $S_1(R)=m_1(R)/m_0(R)$ the factor $g(R)$ cancels out in \eqnref{mp_smooth} as first pointed out in Ref.~\cite{Gus13c}, see also Ref.~\cite{Vos13}.
It is $m_1$ that determines the collision rate. Caustics and clustering contribute additively, Eqs.~(\ref{eq:collision_relvel_moment1}) and \eqnref{Rt0}.
The factorisation (\ref{eq:mp_smooth}) is nevertheless useful at small Stokes numbers, because $m_0$ and
$S_1$ are most easily estimated separately in this limit.

To account for fractal clustering, we use the approximation $m_0(R\ll\eta)\sim R^{d_2-1}$ for small separations.
Here the spatial correlation dimension $d_2$ is related to the phase-space correlation dimension by $d_2=\min(D_2,d)$~\cite{Bec08,Gus11b}.
For large separations we approximate $m_0(R\gg\eta)\sim R^{d-1}$ (uniform distribution).
This is needed to get the normalisation of $m_1(R)$ approximately correct. In flows with an inertial range, a better estimate of $m_0(R)$ is required, or alternatively, the global normalisation could be matched to numerical data.

We match these two asymptotes of $m_0(R)$ at $R=\eta$ to find
\begin{equation}
m_0(R\ll\eta)\approx
dL^{-d}R^{d_2-1}\eta^{d-d_2}\,.
\eqnlab{m0approx}
\end{equation}
We approximate the smooth contribution to $S_1(R)$ as follows.
The relative motion for two particles is obtained by linearisation of \eqnref{eqm_Stokes}.
In random incompressible flows, numerical simulations at $\ku\sim 1$ show that effects due to preferential sampling of the velocity field $\ve u$ are small for all values of $\st$ (in contrast to preferential sampling of flow-velocity gradients).
This allows us to ignore the position dependence in $\ve u(\ve r,t)$.
It is then possible to solve the equation for $V_R$ to obtain~\cite{Gus13c}
\begin{equation}
V_R=\gamma\int_0^t{\rm d}t_1 e^{\gamma(t_1-t)}\Delta\ve u(\ve r_0,\Delta\ve r_0,t_1)\cdot\hat{\ve e}_R(0)\,.
\eqnlab{Vr_ergodic_solution}
\end{equation}
Here $\Delta\ve u$ is the difference of the fluid velocity evaluated at the positions of two particles with  constant positions $\ve r_0$ and $\ve r_0+\Delta\ve r_0$ and $\hat{\ve e}_R(0)$ is the unit vector along their separation.
We remark that provided that preferential sampling can be neglected, \Eqnref{Vr_ergodic_solution} is valid for general values of $\st$. It correctly describes the time lag between the fluid and particle velocities.
We use \Eqnref{Vr_ergodic_solution} to determine the statistics of $V_R$.
In the statistical model $\ve u$ is a Gaussian random function with time-correlation function $\langle u_i(\ve r_0,t)u_j(\ve r_0,0)\rangle\sim\delta_{ij}\exp[-|t|/\tau]$ [see \Eqnref{phiCorr}].
Using \Eqnref{Vr_ergodic_solution} together with the correlation function of $\ve u$ to calculate all moments $\langle V_R^p\rangle$ we find that
the steady-state distribution of $V_R$ conditional on $R$ is Gaussian with variance $\langle(\Delta\ve u\cdot\hat{\ve e}_R)^2\rangle/(1+\st)$.
We use this result to compute $S_1(R)$. Multiplying with \eqnref{m0approx} we find $m_1$ in \eqnref{mp_smooth}. Comparison with \eqnref{collision_relvel_moment1} yields an estimate of $b_1$
(the coefficient of the smooth contribution to $\Rt$ for $D_2\le d$)~\cite{Gus13c}
\begin{equation}
b_1\tau= (\eta/L)^d\ku\sqrt{2d/(\pi(1+\st))}\,.
\eqnlab{collision_collrate_ST_finiteSt}
\end{equation}
The resulting smooth contribution to the collision rate (\ref{eq:Rt0}) is compared to data from numerical simulations for $\ku=0.02$ in \Figref{inertial}{\bf a},
using data for the correlation dimension shown in \Figref{inertial}{\bf b}.
The corresponding comparison for $\ku=1$ is shown in \Figref{inertialKu1}.
When $\st=0$, the collision rate corresponding to \Eqnref{collision_collrate_ST_finiteSt} equals the Saffman-Turner rate $\R_{\rm ST}$.
For finite values of $\st$, the rates $\R_{\rm ST}$ and $\Rt$ differ in two respects.
First, the radial dependence of $\Rt$ scales with the correlation dimension, $a^{d_2}$, as expected in a system with spatial clustering.
This gives a larger contribution than the corresponding factor, $a^{d}$, in $\R_{\rm ST}$.
Second, the reduction of typical relative speeds due to the delay in particle response to changes in the flow, \Eqnref{Vr_ergodic_solution}, gives the factor $(1+\st)^{-1/2}$ in $\Rt$.
This is the main effect for small values of $\st$ (or for large enough values of $a$) in random flows because the correlation dimension $d_2$ scales as $\st^2$ for small values of $\st$.
This is clearly visible for small values of $\ku$ in \Figref{inertial}{\bf a}.
From Figs.~\figref{inertial}{\bf a} and \figref{inertialKu1}{\bf a} it is clear that the smooth part is not the full contribution to the ghost-particle approximation of the collision rate.

\begin{figure*}
\includegraphics[width=14cm]{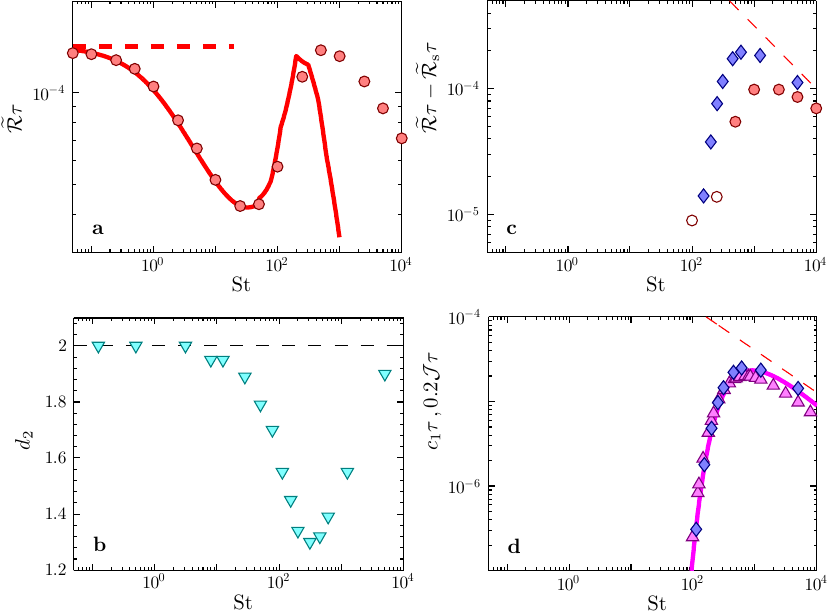}
\caption{\figlab{inertial} {\em (Online colour).}
{\bf a} Same as \Figref{advective}{\bf a} but just $\Rt$
shown. The red solid line shows the
smooth contribution to $\Rt$, \Eqnref{collision_collrate_ST_finiteSt}, using data from panel ${\bf b}$.
{\bf b} Correlation dimension.
Markers show data from numerical simulations (\marker{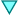}).
{\bf c} Data of panel {\bf a} plotted with the smooth part $\widetilde R_{\rm s}$ given by \Eqnref{collision_collrate_ST_finiteSt} subtracted (\marker{1}). Hollow markers correspond to negative values.
The caustic part of Eqs.~\eqnref{Rt0} and~\eqnref{collision_relvel_moment1} with $c_1$ obtained from the numerical simulations in Ref.~\cite{Gus12}, as described in the text, is plotted as (\marker{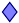}).
The kinetic gas-theory result \eqnref{R_KG} is shown as a dashed line.
{\bf d} $c_1$ from panel {\bf c} compared to \eqnref{c1} with kinetic gas theory (dashed) and $\J$ (\marker{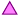}).
$\J$ was evaluated from numerical simulations and plotted with a fitted prefactor $0.2$.
The one-dimensional rate in \Eqnref{J} is plotted as solid magenta with a fitted prefactor $A=0.4$.
Parameters: $\ku=0.02$, $a=0.01\eta$, $N=1000$, $L\approx 5\eta$.
}
\end{figure*}
\begin{figure*}
\includegraphics[width=14cm]{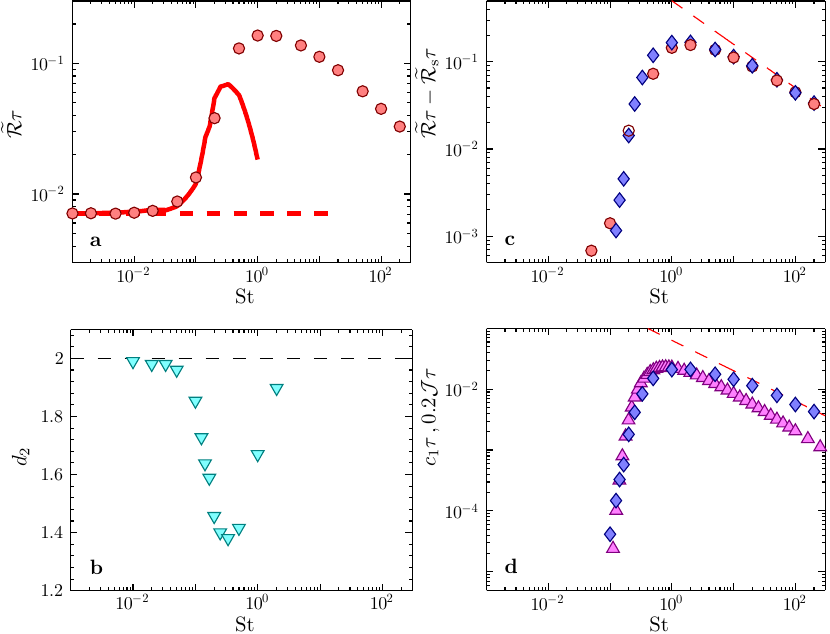}
\caption{\figlab{inertialKu1} {\em (Online colour).}
Same as \Figref{inertial} for $\ku=1$.
The white-noise results, only valid for small values of $\ku$, have been omitted.
The caustic formation rate in panel {\bf d} is plotted with the same prefactor $0.2$ as in \Figref{inertial}.
}
\end{figure*}

\section{Caustic contribution}
\seclab{collision_caustics}

If the $\st$-dependence of the collision rate were determined by clustering only, it would peak for a value of $\st$ close to the value for which the correlation dimension (plotted in Figs.~\figref{inertial}{\bf b} and~\figref{inertialKu1}{\bf b}) has a minimum. Instead, the collision rate increases substantially as $\st$ passes a threshold, and then it decreases only slowly as $\st$ becomes even larger (see Figs.~\figref{inertial}{\bf a} and \figref{inertialKu1}{\bf a}).
In this section we explain this behaviour using \Eqnref{collision_relvel_moment1}.
To isolate the caustic contribution we subtract the smooth contribution \eqnref{collision_collrate_ST_finiteSt} from~$\Rt$.
The result is shown in Figs.~\figref{inertial}{\bf c} and \figref{inertialKu1}{\bf c} (since our expression for $b_1$ and numerical data for~$\Rt$ are only approximate, this subtraction fails when $c_1$ is too small).
The theory from Refs.~\cite{Gus11b,Gus12,Gus13c} does not determine the factor $c_1$.
In Ref.~\cite{Gus12} the coefficient $c_1$ was obtained numerically by fitting (\ref{eq:MWB_ansatz}) to numerical results for the first moment $m_1(R)$ of relative velocities as a function of spatial separation and we use these results for $c_1$ to evaluate \eqnref{Rt0}.
The result is compared to the caustic part of the ghost-particle approximation in \Figref{inertial}{\bf c} for $\ku=0.02$ and in \Figref{inertialKu1}{\bf c} for $\ku=1$.
We find that it agrees well when $\ku=1$. We note that $c_1$ overestimates the caustic contribution to $\Rt$ somewhat when $\ku$ is very small,
due to the fact that the approximation of the smooth part is not precise at intermediate Stokes numbers. But apart from this inaccuracy Figs.~\figref{inertial}{\bf c} and \figref{inertialKu1}{\bf c} demonstrate that the collision rate is a sum of two contributions, a smooth one and a second one due to caustics.

It was argued in Ref.~\cite{Wil06} that $c_1$ is proportional to the caustic formation rate $\J$ for small values of $\st$ (see \Secref{parameterisation}).
A comparison between $c_1$ and $\J$ from numerical simulations is shown in \Figref{inertial}{\bf d} for $\ku=0.02$ and in \Figref{inertialKu1}{\bf d} for $\ku=1$.
We find that both $c_1$ and ${\J}$ show an activated behaviour around the same values of $\st$, but details differ.
This is due to the fact that not all caustics lead to collisions, only caustics with small relative velocities in directions normal to their separation vector, $V^2-V_R^2\ll 1$, may cause collisions between small particles.
The dynamics when $V^2-V_R^2\ll 1$ may to lowest order be approximated by equations for the separation and radial velocity only.
It is therefore a one-dimensional rate ${\J}_{d=1}$ of caustic formation that is relevant for collisions of small particles in $d$ dimensions.
In the white-noise limit ${\J}_{d=1}$ was calculated in Refs.~\cite{Gus13a,Wil03}.
Using a one-dimensional Fokker-Planck description for the distribution $\rho(z)$ of particle velocity gradients, $z\equiv\partial v/\partial x$, and identifying the constant steady-state probability current at which $z$ passes through infinity (a caustic occurs), the caustic formation rate was found to be~\cite{Gus13a,Wil03}
\begin{equation}
{\J}_{d=1}\tau
=\frac{1}{2\pi\st}{\mathcal Im}\Big[\frac{ {\rm Ai}'(y) }{\sqrt{y} {\rm Ai}(y)}\Big]\Bigg|_{y=(-24\,\ku^2\st)^{-2/3}}\,,
\eqnlab{J}
\end{equation}
where ${\rm Ai}$ denotes the Airy function.
At large enough Stokes numbers, particles originating at separations larger than $\eta$ may collide at large relative velocities.
This is not accounted for in the linearized dynamics leading to ${\J}_{d=1}$. Thus $c_1$ cannot be given by ${\J}_{d=1}$  (or the $d$-dimensional rate ${\J}$)
at large Stokes numbers.
We estimate a critical Stokes number where this deviation must occur, using
that the typical relative velocity for separations larger than $\eta$ is $u_0/\sqrt{1+\st}$ in accordance with the smooth relative dynamics discussed in \Secref{collision_smooth}.
For freely moving particles originating at $\eta$ with relative radial speed $u_0/\sqrt{1+\st}$ to collide during the relaxation time $\gamma^{-1}$
we must have $\st>\st_{\rm c}=(1+\sqrt{1+4\ku^2})/(2\ku^2)$.

In the white-noise limit the corresponding value becomes $\st_{\rm c}=1/\ku^2$.
When $\st\gg\st_{\rm c}$, then $\Rt$ must approach the random kinetic gas result, $\R_{\rm g}$~\eqnref{R_KG}.
In summary, we expect
\begin{equation}
c_1\tau \sim \left \{ \begin{array}{ll}
A {\J}_{d=1}\tau & \mbox{if } \st\ll\st_{\rm c} \cr
2(\eta/L)^d\ku\sqrt{d/(\pi\st)} & \mbox{if } \st\gg\st_{\rm c} \cr
\end{array} \right.
\eqnlab{c1}
\end{equation}
In the white-noise limit we find excellent agreement with our numerical results ($A\approx 0.4$), see \Figref{inertial}{\bf d}.

But for $\ku=1$ the picture is more complicated.
As before, the collision rate approaches the kinetic gas limit for $\st\gg\st_{\rm c}$ and there is an activated behaviour for small values of $\st$, but the form of the activation is different from that of the caustic formation rate.
For small values of $\st$, $c_1\sim e^{-S/\st}$ with $S\approx 1$.
As discussed in Ref.~\cite{Gus13a} the caustic formation rate behaves as $e^{-S/\st^2}$ for one-dimensional flows with finite correlation time.
We find from numerical simulations that this is also the case in two spatial dimensions, with $S\approx 0.1$ for $\ku=1$.

Finally we discuss which of the terms in \eqnref{collision_relvel_moment1} dominates. This depends upon the value of $R$ as explained in Ref.~\cite{Gus13c}.
At
$R_{\rm c}=\eta\left(c_1/b_1\right)^{1/(1+D_2-d)}$ the two terms in \eqnref{collision_relvel_moment1} are equal.
$R_{\rm c}$ depends on $\st$ through the parameters $b_1$, $c_1$ and $D_2$.
Evaluating $R_{\rm c}$ for the statistical model used here shows that
$R_{\rm c}$ grows even more rapidly than $c_1$ because of the power $1/(1+D_2-d)$ in $R_{\rm c}$ (c.f. Fig.~9 in Ref.~\cite{Gus13c}).
As soon as caustics are activated, they dominate the collision rate more or less independently of the particle size.
This is consistent with results from direct
numerical simulations of inertial particles in turbulent flows~\cite{Vos13}.

\section{Range of validity}
\Figref{error}{\bf a} shows a comparison of the theory [Eqs.~\eqnref{Rt0} and \eqnref{collision_relvel_moment1}] to numerical simulations of the ghost-particle approximation $\Rt$ for a range of particle sizes and Stokes numbers, and for $\ku=1$.
In \Figref{error}{\bf b} the relative error is plotted.
We observe errors smaller than $10\%$ for most of the data.
The error is significant only when $\st\sim 1$, for particle sizes that are not very small. This is expected, for Stokes numbers of order unity subleading corrections to the asymptotic matching are most important, and the approximation for the smooth part in \Eqnref{collision_collrate_ST_finiteSt} is less accurate.
In \Figref{error}{\bf c} the relative error between the theory and the collision rate~$\R$ is shown.
As expected, when caustics are abundant ($\st$ larger than order unity), the error is approximately the same as for the ghost-particle approximation.
For smaller values of $\st$ the deviation observed in \Figref{advective} shows up.
We observe that this error is approximately independent of the particle size for the range shown in \Figref{error}{{\bf c}}.

\begin{figure*}
\includegraphics[width=18cm]{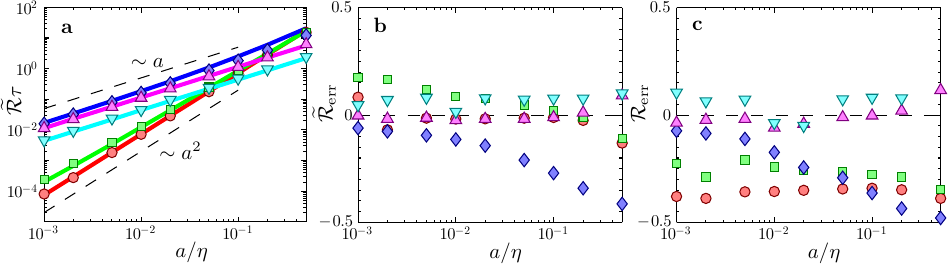}
\caption{\figlab{error} {\em (Online colour).}
{\bf a} Markers show data from numerical simulations of the ghost-particle approximation $\Rt_{\rm data}$ (counting collisions and recollisions) against particle size~$a$.
Lines show the ghost-particle approximation $\Rt_{\rm theory}$ obtained from Eqs.~\eqnref{Rt0} and \eqnref{collision_relvel_moment1} with $b_1$ from \Eqnref{collision_collrate_ST_finiteSt} and $c_1$, $d_2$ according to \Figref{inertialKu1}.
{\bf b} Relative error $\Rt_{\rm err}\equiv\Rt_{\rm data}/\Rt_{\rm theory}-1$ between the markers and lines in panel {\bf a}.
{\bf c} Relative error $\R_{\rm err}\equiv\R_{\rm data}/\Rt_{\rm theory}-1$ between numerical simulations for the collision rate $\R_{\rm data}$ (counting first collisions only, neglecting recollisions) and the the theory $\Rt_{\rm theory}$.
Parameters: $\st=0.01$ (\marker{1}), $\st=0.1$ (\marker{2}), $\st=1$ (\marker{3}), $\st=10$ (\marker{4}), $\st=100$ (\marker{5}), $\ku=1$, $N=1000$, $L\approx 5\eta$.
The data point with $\st=100$ and $a=0.5\eta$ in panel {\bf c} is missing because the particle concentration is too low in the stationary state.
}
\end{figure*}

\section{Conclusions}
Eqs.~(\ref{eq:Rt0},{\ref{eq:collision_relvel_moment1}) determine the collision rate $\Rt$ in the ghost-particle approximation. These equations are obtained using earlier results on relative velocities of inertial particles \cite{Gus11b,Gus12,Gus13c} (see also \cite{Wil06}).
We compared Eqs.~(\ref{eq:Rt0},{\ref{eq:collision_relvel_moment1}) to data from numerical simulations of a statistical model for turbulent aerosols~\cite{Gus15a}.
We found that the ghost-particle approximation may over-count the collision rate for small values of $\st$. But for large values of $\st$,
when caustics are important, this effect is relatively small, see also Ref.~\cite{Vos11}.
$\Rt$ is parameterised by a sum of two contributions, one smooth and one due to caustics. This form is similar but not identical to other parameterisations~\cite{Fal02,Wil06,Duc09,Vos13}.
Our results allow us to determine whether the non-smooth part of the ghost-particle approximation is proportional to the caustic formation rate $\J$.
We find reasonable agreement for not too large values of $\st$, but deviations are clearly observable.
One explanation for this is that only caustics with small relative velocities in directions normal to their separation vector contribute to collisions between small particles.
For large values of $\st$, particles from separations larger than $\eta$ may collide, and $c_1$ is no longer proportional to the locally calculated $\J$.
For the statistical model we estimate the typical scale $\st_{\rm c}$ at which this transition occurs. When $\st\gg \st_{\rm c}$, $c_1$ must approach the kinetic-gas result~\cite{Abr75}.
Our parameterisation allows us to determine when the caustic contribution dominates. We found that it dominates as soon as caustics
are activated, in agreement with the findings in Ref.~\cite{Vos13}.
In comparison of the relative error of the parameterisation and simulation data we found quantitative agreement for the ghost-particle approximation.
For the collision rate we found quantitative agreement when caustics are abundant.

Here we have studied a highly idealised model. In reality the particles collide (and may coalesce) with collision and coalescence efficiencies that
depend on the relative velocities. In the future a much more detailed understanding of the possible outcomes of collisions is required.  To this end it is necessary to not only study the moments of the relative velocities between the inertial particles,
but also their distribution~\cite{Gus13c}. One example in this direction is the study of elastic collisions~\cite{Bec13}.
It would also be interesting to extend the model studied here for particles of different sizes.

{\em Acknowledgements.} Financial support by Veten\-skaps\-r\aa{}\-det [grant number 2013-3992], Formas  [grant number 2014-585], and by the G\"oran Gustafsson Foundation for Research in Natural Sciences and Medicine,
and by the grant \lq Bottlenecks for particle growth in turbulent aerosols\rq{} from the Knut and Alice Wallenberg Foundation, Dnr. KAW 2014.0048 is gratefully acknowledged.
KG also acknowledges funding from the European Research Council under the European Union's Seventh Framework Programme, ERC Grant Agreement No 339032.
The numerical computations were performed using resources provided by C3SE and SNIC.
}

\end{document}